\documentclass[aps,showpacs,preprint]{revtex4}
\usepackage{amstext}
\usepackage{amssymb}
\usepackage{graphicx}

\begin{document}

\title{Energy weighted sums for collective excitations in nuclear Fermi-liquid}
\author{V.M. Kolomietz, S.V. Lukyanov, O.O. Khudenko}
\affiliation{Institute for Nuclear Research, NAS of Ukraine, Prosp. Nauky 47, 03680 Kyiv, Ukraine}


\begin{abstract}
Model independent, $m_1$, adiabatic, $m_{-1}$, and
high-energy, $m_3$, energy weighted sums for the isoscalar and
isovector nuclear excitations are investigated within the
framework of the kinetic theory adopted to the description of a
two-component nuclear Fermi-liquid. For both the adiabatic and
scaling approaches, the connection of the EWS
$m_{-1}$ and $m_3$ to the nuclear stiffness coefficients and the
first- and zero-sound velocity is established. We study the
enhancement factor $\kappa_{I}$ in the energy weighted sum $m'_1$
for the isovector excitations and provide the reasonable
explanation of the experimental exceeding of the 100$\%$
exhaustion of sum $m'_1$ for the isovector giant dipole
resonances. We show the dependence of the enhancement factor
$\kappa_{I}$ on the nuclear mass number $A$ and analyse its
dependence on the Landau's isovector amplitude $F'_1$.
\end{abstract}


\maketitle

\section{Introduction}

The strength function is the basic characteristic which determines
the behavior of a quantum system in an external periodic field
$U_{ext}=\lambda (t)\hat{q}$ \ ($\lambda (t)=\lambda_0e^{-i\omega
t} + c.c.$, here $\hat{q}$ is the transition operator)
\begin{equation}\label{eq1}
S(E)=\sum\limits_{n\ne 0}{\left|{\left\langle{\Psi_n}\right|\,\hat{q}\,\left|
{\Psi_0}\right\rangle}\right|}^2\delta(E-E_n),\quad E=\hbar\omega,
\end{equation}
where $\Psi_n$ and $E_n$ are the eigenfunctions and the
eigenenergies of the total hamiltonian $\hat{H}$, respectively.
Using the strength function $S(E)$ one can calculate the moments
$m_k$ (EWS)
\begin{equation}\label{eq2}
m_k=\int{dE\,S(E)\,E^k}=\sum\limits_{n\ne 0}
{\left|{\left\langle{\Psi_n}\right|\,\hat{q}\,\left|{\Psi_0}
\right\rangle}\right|}^2(E_n-E_0)^k.
\end{equation}
Here, for convenience, we have included the ground state
energy $E_0=0$ into the energy factor. Special role of the EWS
$m_k$ is caused by its connection to the transport characteristic
of the system. For example, the sums $m_{-1}$ and $m_{-3}$
determine the stiffness and mass coefficients for the collective
excitations in the system \cite{BoLaMa}. Determined via the
properties of the ground state of the system the sum $m_1$ plays a
specific role. In many cases, it does not depend on the model used
for the description of the collective motion. This allows one to
test the results of theoretical calculations as well as the
correctness and the completeness of the experimental data.

During a few years, significant attention was paid to the analysis
of EWS for the giant multipole resonances (GMR)
\cite{La,LiBr,HaDi,St,Wo,LiSt,HaSaZh97,HaSaZh98,SaHaZh99}. Nuclear
giant resonances exhaust a significant part of EWS (sometimes near
100{\%}) and establish the relatively simple connection between the
values of $m_k$ and the basic characteristics of the GMR. However,
some problems occur while researching the EWS for the isovector
giant dipole resonances which are the best investigated
experimentally. The connected problem is that the sum $m_1$ is not
model independent because of the dependence of the effective
nuclear forces on the nucleon velocity. Thus, for the theory to
agree with the experimental data, one has to include a phenomenological
enhancement factor to the sum $m_1$ \cite{Wo}. As a
consequence, this leads to the modification of other sums $m_k$
and can affect the definition of the nuclear transport
characteristics.

In this work, we study the EWS $m_k$ for the
isovector collective excitations in heavy nuclei and nuclear
matter. Our approach is based on the kinetic Landau-Vlasov's
theory adopted to a two-component nuclear Fermi-liquid. In
Section 2, we consider the connection between the EWS $m_k$ and the
linear response function. The connection of the Landau's theory of
the Fermi-liquid to the hydrodynamical model and to the scaling
approximation is shown in Section 3 \cite{Ko,KoSh}. In Section 4, we
apply our approach to finite nuclei. The main conclusions
of the work are formulated in Section 5.

\section{Linear response function and EWS}

Let us consider the response of a nucleus on an external
field $U_{ext} (t)$ periodic in time which is switched on
adiabatically at $t=-\infty$:
\begin{equation}\label{eq3}
U_{ext}(t)=\lambda_0 e^{-i(\omega+i0)t}\hat{q}+\lambda_0^\ast
e^{i(\omega-i0)t}\hat{q}^\ast ,
\end{equation}
where $\hat{q}$ is the Hermitian operator,
\begin{equation}\label{eq4}
\hat{q}=\sum\limits_{i = 1}^A {\hat{q}(\vec{r}_i ,\tau_i )},
\end{equation}
$A$ is the mass number, and $\tau_i$ is the isotopic variable. If
$\lambda_0<<1$, then quantum mechanical expectation of the
operator $\hat{q}$ takes the following form (see. \cite{LaLi5})
\begin{equation}\label{eq5}
\left\langle\hat{q}\right\rangle=\chi(\omega)\lambda_0 e^{-i\omega t}
+\chi^\ast(\omega)\lambda_0^\ast e^{i\omega t},
\end{equation}
where $\chi(\omega)$ is the linear response function
\begin{equation}\label{eq6}
\chi(\omega)=\sum\limits_n{\left|{\left\langle{\Psi_n}
\right|\,\hat{q}\,\left|{\Psi_0}\right\rangle}\right|^2}
\left[{\frac{1}{E_n-E_0-\hbar\omega-i0}+\frac{1}{E_n-E_0
+\hbar\omega+i0}}\right].
\end{equation}
Let us introduce the polarization response function
\begin{equation}\label{eq7}
\chi^{(\pi)}(\omega)=Re\chi(\omega)
=-2\sum\limits_n{\left|
{\left\langle{\Psi_n}\right|\,\hat{q}\,\left|{\Psi_0}\right\rangle}
\right|^2\frac{E_n-E_0}{(\hbar\omega)^2-(E_n-E_0)^2}}.
\end{equation}

It is easy to establish the connection between the EWS
$m_k$ and the linear response function $\chi(\omega)$. Let us
take the Taylor expansion of the function $\chi^{(\pi)}(\omega)$
in a series in $\hbar\omega$ as $\omega\to 0$ (adiabatic expansion)
and in a series in $(\hbar\omega)^{-1}$ as $\omega\to\infty$
(high-frequency expansion). Using (\ref{eq2}) and (\ref{eq7}), we
have
\begin{equation}\label{eq8}
\left.{\chi^{(\pi)}(\omega)}\right|_{\omega\to 0}=2\;\left[{m_{-1}
 +(\hbar\omega)^2m_{-3}+...}\right],
\end{equation}

\begin{equation}\label{eq9}
\left.{\chi^{(\pi)}(\omega)}\right|_{\omega\to\infty}=
-\frac{2}{(\hbar\omega)^2}\;\left[{m_1+(\hbar\omega)^{-2}m_3+...}\right].
\end{equation}

Below we will pay a special attention to the investigation of the
sums $m_{-1}$, $m_1$ and $m_3$. Using these sums, one can define
two averaged energies of collective motion
\begin{equation}\label{eq10}
\tilde{E}_1=\sqrt{\frac{m_1}{m_{-1}}} \quad
\textmd{та} \quad \tilde{E}_3=\sqrt{\frac{m_3}{m_1}} .
\end{equation}
It is easy to see that the closeness of the energies $\tilde{E}_1$
and $\tilde{E}_3$ to each other determines the exhaustion of the
EWS $m_k$ by one state $\Psi_n$ (see (\ref{eq2})). If the
effective nuclear forces do not depend on the nucleon velocity,
then the sum $m_1$ can be easily calculated and takes the form
which does not depend on the model of collective motion. Namely,
\begin{equation}\label{eq11}
m_1=\frac{1}{2}\left\langle{\Psi_0\left|{\left[{\hat{q},\left[{\hat{q},
\hat{H}}\right]}\right]}\right|\Psi_0}\right\rangle
=\sum\limits_{n\ne 0}{\left|{\left\langle{\Psi_n }
\right|\,\hat{q}\,\left|{\Psi_0}\right\rangle}\right|}^2(E_n-E_0)
=\frac{\hbar^2}{2m}\int{d\vec{r}}\,\rho_{eq}(\vec{r})
\vert\vec{\nabla}\hat{q}(\vec{r})\vert^2,
\end{equation}
where $\rho_{eq}(\vec{r})$ is the nucleon density for the ground
state of the nucleus
\[
\rho_{eq}(\vec{r})=\left\langle{\Psi_0\left|{\sum\limits_{i=1}^A
{\delta(\vec{r}-\vec{r}_i)}}\right|\Psi_0}\right\rangle.
\]
(Here, and in the following, the symbol "eq" means that the proper
value is related to the equilibrium (basic) state of the nucleus.)
The expression (\ref{eq11}) is the so-called model independent
EWS rule. If only one (collective) state
$\Psi_{n=G}$ exhausts the sum rule (\ref{eq11}), i.e.,
\begin{equation}\label{eq12}
m_1\approx\left|{\left\langle{\Psi_G}\right|\,\hat{q}\,\left|{\Psi_0}
\right\rangle}\right|^2(E_G-E_0),
\end{equation}
then we have $\tilde{E}_1\approx\tilde{E}_3$ from (\ref{eq2}),
(\ref{eq10}).

The low-frequency (adiabatic) sum $m_{-1}$ is connected to the
nuclear stiffness under the adiabatic slow deformation of the
nucleus, in another words under the deformation that does not lead to
the quantum transitions between nuclear levels. To reveal this
connection, we will evaluate the energy variation $\Delta E$ of
the nuclear ground state in an external \textit{static} field
$U_{ext}=\lambda_0\hat{q}$ for $\lambda_0\to 0$. Using the quantum
perturbation theory for the calculation of the wave function
$\Psi$ of Hamiltonian ${\hat{H}}'=\hat{H}+\lambda_0\hat{q}$ in
the second order in the small parameter $\lambda_0$, we obtain
\begin{equation}\label{eq13}
\Delta E_{ad}=\left\langle\Psi\right|\,\hat{H}\,\left|\Psi
\right\rangle-\left\langle{\Psi_0}\right|\,\hat{H}\,\left|{\Psi_0}
\right\rangle=\lambda_0^2 \,m_{-1}.
\end{equation}
Let us calculate the variation of the nuclear form parameter
$Q=\left\langle{\Psi\vert\hat{q}\vert\Psi}\right\rangle$ in the
external field $\lambda_0\hat{q}$,
\begin{equation}\label{eq14}
\Delta Q=Q=\left\langle\Psi\right|\,\hat{q}\,\left|\Psi
\right\rangle-\left\langle{\Psi_0}\right|\,\hat{q}\,\left|{\Psi_0}
\right\rangle=2\,\lambda_0m_{-1},
\end{equation}
where we have assumed
$\left\langle{\Psi_0}\right|\,\hat{q}\,\left|{\Psi_0}\right\rangle=0$.
From (\ref{eq13}) and (\ref{eq14}), we find the nuclear stiffness
parameter $C_{Q,ad}$ with respect to the adiabatic change of the
nuclear form as
\begin{equation}\label{eq15}
C_{Q,ad}=\frac{\partial^2\Delta E_{ad}}{\partial Q^2}=\frac{1}{2\,m_{-1}}.
\end{equation}

Let us now consider the high-frequency sum $m_3$. We
introduce the wave function $\Psi_{sc}$, which is obtained from
the wave function of the nuclear ground state $\Psi_0$ by means of
the scale transformation (scaling-approach),
\begin{equation}\label{eq16}
\Psi_{sc}=e^{\nu [\hat{H},\hat{q}]}\Psi_0 ,
\end{equation}
where $\nu$ is the small parameter of the scale transformation.

In the case of a many-particle wave function $\Psi_0$ given by the
determinant built on the one-particle wave functions
$\phi_\alpha(\vec{r})$, the exponential operator of the scale
transformation in (\ref{eq16}) acts on each function $\phi_\alpha
(\vec{r})$ independently. For example, at the quadrupole
deformation
\[
\hat{q}=\sum\limits_{i=1}^A (r_i^2-3z_i^2),
\]
one can see from (\ref{eq16}) that $\Psi_{sc}$ is also a
determinant which is built on the  one-particle functions
$\phi_{\alpha,sc}(\vec{r})$ obtained by the scale transformation
of coordinates. Namely,
\begin{equation} \label{eq17}
\phi_{\alpha,sc}(\vec{r})\equiv\phi_{\alpha,sc} (x,y,z)
=\phi_\alpha (e^{\tilde{\nu}}x,\,\,e^{\tilde{\nu}}y,\,\,
e^{-2\tilde{\nu}}z),
\end{equation}
where $\tilde{\nu}=-2\hbar^2\nu /m$. As can be seen from
(\ref{eq17}), the scale transformation does not violate the
orthonormalization of the wave functions. By means of
(\ref{eq16}), the energy change $\Delta E$ can be found within the
scaling approximation as
\begin{equation}\label{eq18}
\Delta E=\left\langle{\Psi_{sc}}\right|\,\hat{H}\,\left|{\Psi_{sc}}
\right\rangle-\left\langle{\Psi_0}\right|\,\hat{H}\,\left|{\Psi_0}
\right\rangle=\nu^2\,m_3.
\end{equation}
Using (\ref{eq16}), we obtain the connection between the parameter
of scale transformation $\nu$ and the deformation parameter $Q$:
\begin{equation}\label{eq19}
Q=\left\langle{\Psi_{sc}}\right|\,\hat{q}\,\left|{\Psi_{sc}}
\right\rangle-\left\langle{\Psi_0}\right|\,\hat{q}\,\left|{\Psi_0}
\right\rangle=2\,\nu\,m_1.
\end{equation}
Finally, using (\ref{eq18}) and (\ref{eq19}), we obtain the
nuclear stiffness coefficient $C_{Q,sc}$ in scaling approximation as
\begin{equation}\label{eq20}
C_{Q,sc}=\frac{\partial^2\Delta E}{\partial q^2}=\frac{m_3}{2\,m_1^2},
\end{equation}
which differs significantly from the adiabatic one $C_{Q,ad}$
(\ref{eq15}). The reasons of such a deference will be made clear in
the next section.

\section{Response function and EWS for nuclear Fermi-liquid}

It is necessary to make some additional assumptions for the
practical calculation of the linear response function
$\chi(\omega)$ and the corresponding EWS $m_k$. We will restrict
ourselves to the Landau's approximation for a nuclear
Fermi-liquid and use the linearized Landau-Vlasov equation
\cite{AbKha}. In the two-component nuclear Fermi-liquid, it is
necessary to consider two possibilities: isoscalar excitations
(when protons and neutrons move in phase) and isovector
excitations (when protons and neutrons move in antiphase).
\bigskip

\subsection{Isoscalar excitations}

For the nuclear matter in a volume $V$ in the case of isoscalar
excitations, the linearized kinetic Landau-Vlasov equation has the
same form as that for a one-component Fermi-liquid \cite{KoSh}
\begin{equation}\label{eq21}
\frac{\partial}{\partial t}\delta f+\vec{v}\cdot\vec{\nabla}_r
\delta f-\vec{\nabla}_p f_{eq}\cdot\vec{\nabla}_r(\delta U_{self}
+ U_{ext})=0,
\end{equation}
where $\delta f=\delta f_n+\delta f_p\equiv\delta
f(\vec{r},\vec{p};t)$ is the variation of the nucleon distribution
($\delta f_n$ for neutrons and $\delta f_p$ for protons) in a
phase space, $\vec{v}$ is the nucleon velocity,
$f_{eq}=f_{eq,n}+f_{eq,p}\equiv f_{eq}(\vec{r},\vec{p})$ is the
equilibrium distribution function, $\delta U_{self}\equiv\delta
U_{self}(\vec{r},\vec{p};t)$ is a variation of the self-consistent
mean field. The subscripts at $\vec{\nabla}$ in
(\ref{eq21}) indicate the variables of differentiation. The
variation of the self-consistent field $\delta U_{self}$ depends
on the effective nucleon-nucleon interaction $v_{int}$. In the
case of homogeneous nuclear matter it is given by
\begin{equation}\label{eq22}
\delta U_{self}=\int\frac{2Vd\vec{p}'}{(2\pi\hbar)^3} \
v_{int}(\vec{p},\vec{p}')\ \delta f(\vec{r},\vec{p}';t),
\end{equation}
where the additional factor 2 at the numerator is due to the spin
degeneration.

The effective interaction $v_{int} (\vec{p},{\vec {p}}')$ is
connected to the Landau's interaction amplitudes $F_l$ \cite{KoSh}
\begin{equation}\label{eq23}
v_{int} (\vec{p},\vec{p}')=\frac{1}{N_F}\sum\limits_{l=0}^\infty
{F_l} P_l (\cos\theta_{p{p}'}).
\end{equation}
Here, $P_l(x)$ are the Legendre polynomials, $\theta_{pp'}$ is the
angle between the vectors $\vec{p}$ and $\vec{p}'$ and $N_F$ is the
density of states near the Fermi surface,
\begin{equation}\label{eq24}
N_F=-4\pi\int\frac{2Vp^2}{(2\pi\hbar )^3} \
\frac{\partial f_{eq}}{\partial\varepsilon_p} \ dp =
\frac{Vm^*p_F}{\pi^2\hbar^3},
\end{equation}
where $\varepsilon_p = p^2/2m^*$, $m^*$ is the effective mass of
a nucleon (the definition of $m^*$ is given below), and $p_F$ is the
Fermi momentum. In (\ref{eq24}), we have used the equilibrium
Fermi distribution function $f_{eq}=\theta
(\varepsilon_F-\varepsilon_p)$, where $\theta (x)$ is the
Heaviside step function and $\varepsilon_F=p_F^2/2m^*$ is the
Fermi energy. The presence of components with $\ell\ne 0$ in sum
(\ref{eq23}) caused by the dependence of the nuclear forces on the
nucleon velocities. Below, we will restrict ourselves to the most
important case where
\begin{equation}\label{eq25}
F_0\ne 0,\quad F_1\ne 0,\quad F_{l\ge 2}= 0.
\end{equation}
Note, that the interaction amplitude $F_1$ determines the
effective mass of a nucleon \cite{AbKha}
\begin{equation}\label{eq26}
m^*=(1+F_1/3)m.
\end{equation}

We will introduce a variation of the nucleon density
$\delta\rho\equiv\delta\rho (\vec{r},t)$ and the isoscalar
velocity field $\vec{u}=\vec{u}(\vec{r},t)$, which are connected
to a variation of the distribution function $\delta
f=f-f_{eq}\equiv\delta f(\vec{r},\vec{p};t)$ by the relations
\begin{equation}\label{eq27}
\delta\rho =\int{\frac{2\,d\vec{p}}{(2\pi\hbar )^3}\delta f,} \quad
\vec{u}=\frac{1}{\rho}\int{\frac{2\,d\vec{p}}{(2\pi\hbar )^3}
\frac{\vec{p}}{m}\delta f \approx}\frac{1}{\rho_{eq}}\int
{\frac{2\,d\vec{p}}{(2\pi\hbar)^3}\frac{\vec{p}}{m}\delta f},
\end{equation}
where
\begin{equation}\label{eq28}
\rho\equiv\rho (\vec{r},t)=\int{\frac{2\,d\vec{p}}{(2\pi\hbar)^3}f}
(\vec{r},\vec{p};t), \quad
\rho_{eq}\equiv\rho_{eq}(\vec{r})=\int{\frac{2\,d\vec{p}}{(2\pi\hbar)^3}
f_{eq}}(\vec{r},\vec{p})
\end{equation}
is the nucleon density. The velocity field $\vec{u}$ and the
variation of the nucleon density $\delta\rho$ satisfy the
continuity relation
\begin{equation}\label{eq29}
\frac{\partial}{\partial t}\delta\rho +\vec{\nabla}\rho \,\vec{u}=0.
\end{equation}
To check this relation, we will calculate the zero-moment of the
kinetic equation (\ref{eq21}). Multiplying Eq. (\ref{eq21}) by
$2d\vec{p}/(2\pi\hbar)^3$ and integrating over $\vec{p}$, we obtain
\begin{equation} \label{eq30}
\frac{\partial}{\partial t}\delta\rho+\vec{\nabla}_r\frac{m}{m^*}
\rho \,\vec{u}+\int{\frac{2d\vec{p}}{(2\pi\hbar)^3}}f_{eq}\vec{\nabla}_r
\cdot\vec{\nabla}_p\delta U_{self}=0.
\end{equation}
Using Eqs. (\ref{eq23})-(\ref{eq25}) and (\ref{eq28}), we have
\begin{equation}\label{eq31}
\delta U_{self}=\frac{V}{N_F}\left({F_0\,\delta\rho+\frac{F_1}{p_F^2}
m\rho\,\vec{p}\cdot\vec{u}}\right).
\end{equation}
Substituting Eq. (\ref{eq31}) into Eq. (\ref{eq30}) and taking
the definition of $m^*$ (\ref{eq26}) into account, we derive the
continuity equation (\ref{eq29}).

To solve the kinetic equation (\ref{eq21}), we assume that
the external field is given by a plane wave $\lambda_0
e^{i(\vec{q}\cdot\vec{r}-\omega t)}$. Then the solution of Eq.
(\ref{eq21}) can be presented as \cite{AbKha}
\begin{equation}\label{eq32}
\delta f\equiv\delta f_{\vec{q}}(\vec{r},\vec{p};t)=
-\frac{\partial f_{eq}}{\partial\varepsilon_p}\nu_{\vec{q}}
(\vec{p})\,e^{i(\vec{q}\cdot\vec{r}-\omega t)},
\end{equation}
where $\nu_{\vec{q}}(\vec{p})$ is the unknown function.
Substituting Eq. (\ref{eq32}) into Eq. (\ref{eq21}), we obtain the
following equation for $\nu_{\vec{q}}(\vec{p})$
\begin{equation} \label{eq33}
(\omega-\vec{q}\cdot\vec{v})\,\nu_{\vec{q}}(\vec{p})+\vec{q}
\cdot\vec{v}\int{\frac{2Vd{\vec{p}}'}{(2\pi\hbar )^3}\,}v_{int}
(\vec{p},{\vec{p}}')\,\,\frac{\partial f_{eq}}{\partial
\varepsilon_{p}'} \nu_{\vec{q}}(\vec{p})+\lambda_0 \,\vec{q}\cdot\vec{v}=0.
\end{equation}
Let us expand the function $\nu_{\vec{q}}(\vec{p},t)$ in a
power series in the multipolarity $l$ of a Fermi surface
distortion
\begin{equation}\label{eq34}
\nu_{\vec{q}}(\vec{p})=\sum\limits_{l = 0}^\infty \,P_l
(\cos\theta_{pq})\,\nu_l,
\end{equation}
where $\theta_{pq}$ is the angle between the vectors $\vec{p}$ and
$\vec{q}$. Using Eqs. (\ref{eq23}), (\ref{eq24}), (\ref{eq34}) and
(\ref{eq33}), we obtain the infinite set of equations for the
amplitudes $\nu_l$ \cite{BaPe}:
\begin{equation}\label{eq35}
\nu_l+(2l+1)\sum\limits_{l'}\frac{Q_{ll'}(s)}{2{l}'+1}
F_{l'}\nu_{l'} -\lambda_0(2l+1)Q_{l0}(s)=0.
\end{equation}
Here, $s=\omega /qv_F$ and
\begin{equation}\label{eq36}
Q_{ll'}(s)=\frac{1}{2}\int\limits_{-1}^1 {dx\,P_l(x)\frac{x}{x-s}P_{l'}(x)}.
\end{equation}

With regard for condition (\ref{eq25}), (\ref{eq35}) yields
\begin{equation}\label{eq37}
\nu_0(s)=\frac{Q_{00}(s)(1+F_1/3)}{1+F_1/3+Q_{00}(s)(F_0+F_0F_1/3+F_1
s^2)}\lambda_0,
\end{equation}
where we have used the relations \cite{BaPe}
\begin{equation}\label{eq38}
Q_{10}(s)=s\,Q_{00}(s),\quad\quad Q_{11}(s)=s\,Q_{10}(s)+\frac{1}{3}.
\end{equation}
The Legendre function of the second kind $Q_{00}(s)$ can be
calculated by the use of Eq. (\ref{eq36}). Taking the additional
condition of analytical extension of $Q_{00}(s)$ into the complex
plane $s$ into account, we can represent the function $Q_{00}(s)$ as
\begin{equation}\label{eq39}
Q_{00}(s)=1+\frac{s}{2}\,\ln\left|{\frac{s-1}{s+1}}\right|+i\frac{\pi}{2}
s\,\theta(1-\vert s\vert).
\end{equation}

Let us evaluate the density-density response function assuming
$\hat{q}=e^{-i\vec{q}\cdot\vec{r}}$ in Eqs.
(\ref{eq3})-(\ref{eq5}). Using the definition of the linear
response function $\chi(\omega)$ from (\ref{eq5}) and the
relations (\ref{eq27}), (\ref{eq32}), (\ref{eq34}), we obtain
\begin{equation}\label{eq40}
\chi(\omega)=\frac{\left\langle e^{-i\vec{q}\cdot\vec{r}}\right\rangle}
{\lambda_0 e^{-i\omega t}}
=\frac{1}{\lambda_0 e^{-i\omega t}}\int d\vec{r}\int\frac{2d\vec{p}}{(2\pi\hbar )^3}
e^{-i\vec{q}\cdot\vec{r}}\delta f(\vec{r},\vec{p};t)
=\frac{1}{\lambda_0 }N_F\nu_0 (s).
\end{equation}

Finally, taking (\ref{eq37}) into account, we obtain the
density-density response function as
\begin{equation}\label{eq41}
\chi(\omega)=\frac{\overline{Q}_{00}(s)}{1-\kappa(s)\overline{Q}_{00}(s)}
\end{equation}
where
\[
\kappa (s)=-\frac{1}{N_F}\left( {F_0+\frac{F_1}{1+F_1/3}s^2}\right),
\quad \overline{Q}_{00}(s) = N_F Q_{00}(s).
\]
Function (\ref{eq41}) has the same form as the
collective linear response function in the general theory of
collective motion (see e.g., \cite{BoMo}). The quantity
$\overline{Q}_{00}(s)$ is the intrinsic response function, and
$\kappa(s)$ plays the role of the effective interaction parameter.

\begin{figure}
\includegraphics[width=8cm]{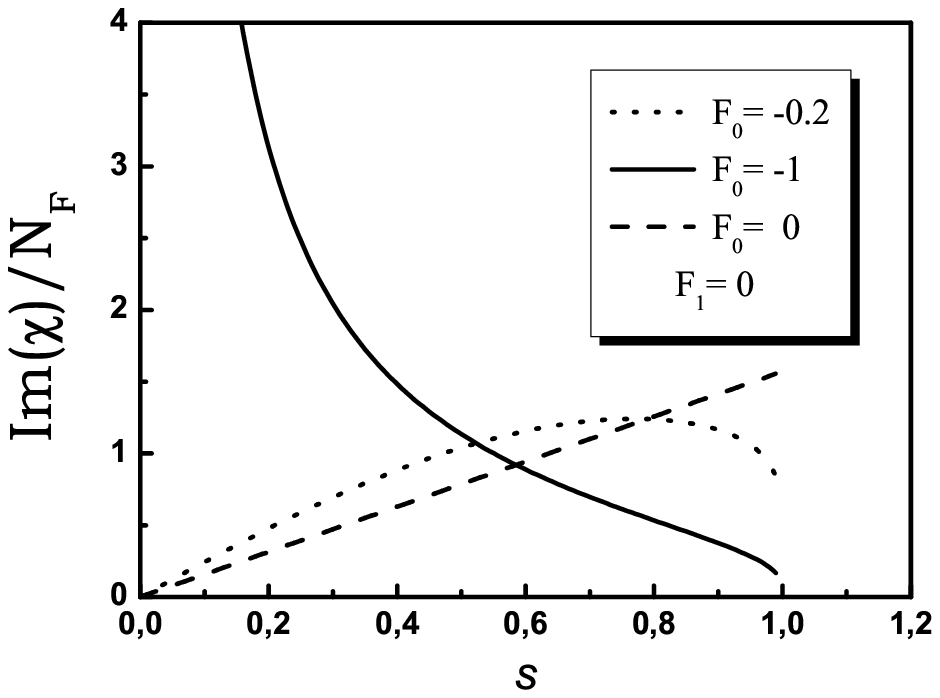}\includegraphics[width=8cm]{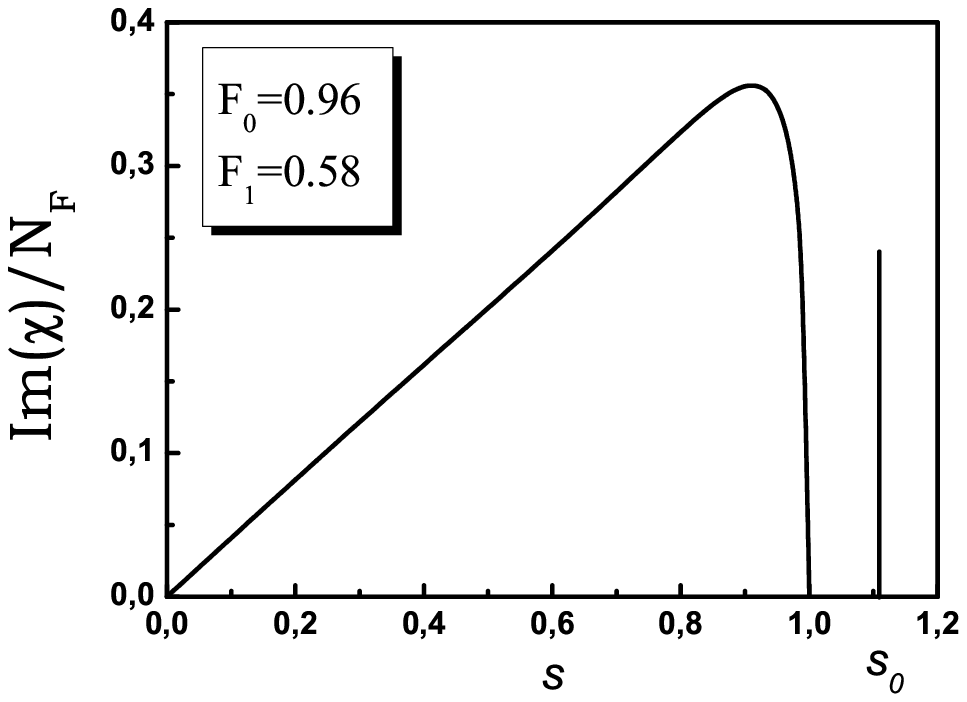}
\caption{Dependences of the strength function $\mathrm{Im}\chi(s)$ on
the dimensionless parameter $s$ for isoscalar excitations: the
left panel is for the Landau damping regime, $-1<F_0\le 0$; the right
panel is for the zero-sound regime, $F_0>0$.}
\label{fig1}
\end{figure}

In \figurename\  \ref{fig1}, we present the dissipative response function
\begin{equation}\label{eq42}
\chi^{(d)}(\omega)=\mathrm{Im}\chi(\omega),
\end{equation}
which is obtained from Eq. (\ref{eq41}) for two regimes: the
Landau damping regime $-1<F_0<0$, left panel, and the zero-sound
regime $F_0>0$, right panel.
Note that the zero-sound mode is dumped at $s<1$ (the Landau
damping \cite{LaLi10}). Here, the zero-sound wave propagates in
phase with some particles and the energy transfer averaged over time
from the wave to particles can be positive. The non-dumped
sound wave exists in a Fermi liquid under the assumption $s>1$ only.
The dimensionless velocity of a sound wave $s$ is determined by the
Landau's dispersion equation \cite{AbKha}
\begin{equation}\label{eq43}
1-\kappa(s)\overline{Q}_{00}(s)=0.
\end{equation}
If the dispersion equation (\ref{eq43}) is satisfied, both the
response function (\ref{eq41}) and the sound wave amplitude
grow to infinity, and the sound wave propagation cannot be
described within the framework of a linear response theory. The
analysis showed \cite{AbKha,LaLi10} that the solution to
Eq. (\ref{eq43}) exists (for real values of $s$) at $F_0>0$
only. This is illustrated in \figurename\  \ref{fig1}. As can be
seen from the right panel of \figurename\ \ref{fig1}, the isolated
root of Eq. (\ref{eq43}) exists at $s>1$ (zero-sound) for $F_0>0$
only.

It is easy to see from the dispersing equation (\ref{eq43}) that
the velocity of the zero-sound increases monotonically with the
interaction parameter $F_0$.

It is useful to consider the solution of the dispersion equation
(\ref{eq43}) at the asymptotic regime $s\to\infty$ (or
$F_0\to\infty)$. Let us use the asymptotic expansion of the
Legendre function of the second kind
\begin{equation}\label{eq44}
\left.{Q_{00}(s)}\right|_{s\to\infty}=-\frac{1}{3s^2}-
\frac{1}{5s^4}-\frac{1}{7s^6}- ... .
\end{equation}
From relations (\ref{eq43}) and (\ref{eq44}), we find the velocity
of the zero-sound wave $u_0=s\,v_F$ at $F_0\to\infty$:
\begin{equation}\label{eq45}
\left.{u_0}\right|_{F_0\to\infty}=\;\,\left. s
\right|_{F_0\to\infty}v_F=\sqrt{\frac{F_0}{3mm^*}p_F^2} .
\end{equation}
Formula (\ref{eq45}) can be compared with that for the velocity
$u_1$ of the normal sound (first-sound) in a classical liquid
\[
u_1=\sqrt{\frac{K}{9m}} ,
\]
where $K$ is the incompressibility coefficient. For the
Fermi-liquid, the incompressibility coefficient is given by
\cite{KoSh}
\begin{equation}\label{eq46}
K=6\,\frac{p_F^2}{2m^*}\,(1+F_0)\approx 220\,MeV
\end{equation}
and we obtain
\begin{equation}\label{eq47}
u_1=\sqrt{\frac{K}{9m}}=\sqrt{\frac{1+F_0}{3mm^*}p_F^2}.
\end{equation}
Taking Eqs. (\ref{eq45})-(\ref{eq47}) into account, we derive
\begin{equation}\label{eq48}
\left.{u_0}\right|_{F_0\to\infty}=\left.{u_1}\right|_{F_0\to\infty}
\end{equation}
This result means that the velocities of the zero- and first
sounds in Fermi-liquid coincide at a significant, $F_0>>1$,
repulsion between the particles.

Using the expansion of the polarization response function
$\chi^{(\pi)}(\omega)=\mathrm{Re}\chi(\omega)$ (\ref{eq8}),
(\ref{eq9}) and expression (\ref{eq41}), one can find the
EWS $m_{-1}$, $m_1$ and $m_3$ for the
Fermi-liquid (see also \cite{LiSt}) as
\begin{equation}\label{eq49}
m_{-1}=\frac{A}{2}\frac{9}{K}, \quad
m_1=\hbar^2\frac{A}{2m}\,q^2, \quad
m_3=\hbar^4\frac{A}{2}\frac{{K}'}{9m^2}\,q^4.
\end{equation}
Here, we have introduced the renormalized (due to the Fermi
surface distortion effect) incompressibility coefficient
${K}'=K+24\varepsilon_F/5$, see \cite{Ko}. Using relations
(\ref{eq10}) and (\ref{eq49}), we can derive the average excitation
energy (the energy centroids of giant isoscalar resonances) in the
adiabatic, $\tilde{E}_1$, and scaling $\tilde{E}_3$, approximations:
\begin{equation}\label{eq50}
\quad \tilde{E}_1=\hbar\sqrt{\frac{K}{9m}} \,q,\quad \quad
\tilde{E}_3=\hbar\sqrt{\frac{{K}'}{9m}} \,q.
\end{equation}
Using the dispersion relation $\tilde{E}=\hbar\tilde{u}\,q$
between the excitation energy of a sound wave, $\tilde{E}$, and the
sound velocity, $\tilde{u}$, and applying Eqs. (\ref{eq50}) and
(\ref{eq26}), we obtain the sound velocity in the adiabatic,
$\tilde{u}_1$, and scaling, $\tilde{u}_3$, approximations:
\begin{equation}\label{eq51}
\tilde{u}_1=\sqrt{\frac{(1+F_0)p_F^2}{3mm^*}} \,,\quad \quad
\tilde{u}_3=\sqrt{\frac{(9/5+F_0)p_F^2}{3mm^*}} \,.
\end{equation}
By comparing Eq. (\ref{eq47}) and Eq. (\ref{eq51}), it can
be seen that the sound velocity in the adiabatic approximation,
$\tilde{u}_1$, coincides with the first sound one, $u_1$, and that
the sound velocity in the scaling approach, $\tilde{u}_3$, exceeds
$u_1$ significantly. The origin of this effect is the same as in
the case of the nuclear stiffness coefficients $C_{Q,ad}$ and
$C_{Q,sc}$, see Eqs. (\ref{eq15}) and (\ref{eq20}).

To clarify the nature of this effect, we will return to the
kinetic equation (\ref{eq33}) and consider the recurrence method
of its solution. For a simplification, we neglect the external
field in (\ref{eq33}) assuming $\lambda_0=0$ and use, instead of
(\ref{eq34}), the following expansion of the amplitude
$\nu_{\vec{q}}(\vec{p},t)$ into a series in the multipolarity $l$
of a dynamic Fermi surface distortion:
\begin{equation}\label{eq52}
\nu_{\vec{q}}(\vec{p})=\sum\limits_{l=0}^\infty
\,\sum\limits_{m=-l}^l{\nu_{lm}(q)} \,Y_{lm}(\hat{p}).
\end{equation}
Substituting amplitude (\ref{eq52}) into Eq.
(\ref{eq33}), using the expressions (\ref{eq23}) and (\ref{eq24}),
multiplying then Eq. (\ref{eq33}) by the spherical function
$Y_{lm}^*(\hat{\vec{p}})$, and integrating over the angles of the
unit vector $\hat{\vec{p}}=\vec{p}/p$, we obtain the following
equation for amplitudes $\nu_{lm}$:
\begin{equation}\label{eq53}
\omega\nu_{lm}-v_F q\,\sum\limits_{{l}'{m}'}{G_{l}' \,\left\langle
{lm\left|{\hat{\vec{q}}\cdot\hat{\vec{p}}}\right|{l}'{m}'}
\right\rangle \,}\nu_{{l}'{m}'}=0.
\end{equation}
Here $\hat{\vec{q}}=\vec{q}/q$, $G_l=1+F_l/(2l+1)$,
\[
\left\langle{lm\left|{\hat{\vec{q}}\cdot\hat{\vec{p}}}
\right|{l}'{m}'}\right\rangle
\equiv C(lm,{l}'{m}')=\int{d\Omega_{\vec{p}}}Y_{lm}^*
(\hat{\vec{p}})\cos\theta_{qp} Y_{{l}'{m}'}(\hat{\vec{p}})=
\]
\begin{equation}\label{eq54}
=(-1)^m\frac{\sqrt{(2l+1)(2{l}'+1)}}{3}\left\langle{l{l}'00\vert
10}\right\rangle
\left\langle{l{l}'m,-{m}'\vert 1,m-{m}'}
\right\rangle,
\end{equation}
where $\left\langle{l_1 l_2 m_1 m_2 \vert lm}\right\rangle$ are
the Clebsh-Gordon coefficients. We will restrict ourselves to the
longitudinal sound waves with $\nu_{l,m\ne 0}=0$ \cite{LaLi10}.
Taking condition (\ref{eq25}) for $\nu_{l0}$ into account, we
obtain the following chain of recurrence equations from (\ref{eq53}):
\begin{eqnarray}
s\nu_{00}-\frac{1}{\sqrt 3}G_1\nu_{10}=0, \nonumber \\
s\nu_{10}-\frac{1}{\sqrt 3}G_0\nu_{00}-\frac{2}{\sqrt{15}}G_2 \nu_{20}=0, \nonumber \\
s\nu_{20}-\frac{2}{15}G_1\nu_{10}-\frac{3}{\sqrt{35}}G_3\nu_{30}=0, \nonumber \\
.\ .\ .\ .\ .\ .\ .\ .\ .\ .\ .\ .\ .\ .\ .\ .\ .\ .\ .\ .\ .\ .\ .\ .\ .\ .\ .\ .\ .\  \label{eq55} \\
s\nu_{l0}-\frac{1}{3}\sqrt{4l^2-1}\left|{\left\langle{ll-100\vert 10}\right\rangle}\right|^2\nu_{l-1,0}
-\frac{1}{3}\sqrt{(2l+1)(2l+3)} \,\left|{\left\langle{ll+100\vert 10}
\right\rangle}\right|^2\nu_{l+1,0}=0, \nonumber
\end{eqnarray}
Under some additional assumptions, the infinite chain of Eqs.
(\ref{eq55}) can be cat-off to obtain the analytical solution.
We will consider two important cases. (i) Neglecting the Fermi
surface distortions with multipolarity $l\geq2$ in Eqs.
(\ref{eq55}), we obtain the solution
\begin{equation}\label{eq56}
\omega=\frac{1}{\sqrt{3}}v_F q\sqrt{G_0 G_1}.
\end{equation}
Consequently, the sound speed is given by
\begin{equation}\label{eq57}
u=\omega/q=\frac{1}{\sqrt{3}}v_F\sqrt{G_0
G_1}=\sqrt{\frac{(1+F_0)p_F^2}{3mm^*}}.
\end{equation}
This result coincides with that for the first sound velocity
$u_1$ of Eq. (\ref{eq47}). Thus, the first sound regime
corresponds to the excitations which preserve the spherical
symmetry of the Fermi surface and leads to a displacement of the
Fermi sphere as a whole. (ii) If we consider
three first equations in (\ref{eq55}) and neglect the Fermi
surface distortions with multipolarity $l\geq3$, then the solution
to Eqs. (\ref{eq55}) (now closed) gives the eigenfrequency
\begin{equation}\label{eq58}
\omega=\frac{1}{\sqrt{3}}v_F q\sqrt{(G_0+4/5)G_1}
\end{equation}
and the sound velocity
\begin{equation}\label{eq59}
u=\omega/q=\frac{v_F}{\sqrt{3}}\sqrt{(G_0+4/5)G_1}=
\sqrt{\frac{(9/5+F_0)p_F^2}{3mm^*}}.
\end{equation}
The sound velocity given by Eq. (\ref{eq59}) coincides with
$\tilde{u}_3$, obtained in the scaling approximation (\ref{eq51}).
Thus, the scaling approximation for a Fermi-liquid means that
all lower multipolarities of a Fermi surface distortion up to
$l=2$ are taken into account. As can be seen from (\ref{eq49}),
the model independent sum $m_1$, as it should be, does not depend
on the nuclear interaction (Landau's amplitudes $F_l$). However,
the last statement is not correct in the case of specific nuclear
excitations, where the sound wave occurs due to the antiphase
motion of the neutrons and the protons (isovector vibrations).

\subsection{Isovector excitations}

Below we consider the isovector excitations when protons and
neutrons move in antiphase. In this case, we rewrite the
kinetic equation (\ref{eq21}) for the protons and the neutrons
separately:
\begin{equation}\label{eq60}
\frac{\partial}{\partial t}\delta f_p+\vec{v}\cdot\vec{\nabla}_r
\delta f_p-\vec{\nabla}_p f_{p,eq}\cdot\vec{\nabla}_r(\delta U_{p,self}
+ U_{p,ext})=0,
\end{equation}
\begin{equation}\label{eq61}
\frac{\partial}{\partial t}\delta f_n+\vec{v}\cdot\vec{\nabla}_r
\delta f_n-\vec{\nabla}_p f_{n,eq}\cdot\vec {\nabla }_r (\delta
U_{n,self} + U_{n,ext} ) = 0.
\end{equation}
We neglect the Coulomb interaction and assume $N=Z$. The corresponding
corrections are not important on the description of the main
characteristics of isovector giant resonances. Subtracting Eq.
(\ref{eq60}) from Eq. (\ref{eq61}) and introducing an isovector
variation of the distribution function
\begin{equation}\label{eq62}
\delta{f}'=\delta f_n-\delta f_p,
\end{equation}
we obtain the kinetic equation for the isovector excitations
\begin{equation}\label{eq63}
\frac{\partial}{\partial t}\delta{f}'+\vec{v}\cdot\vec{\nabla}_r
\delta{f}'-\vec{\nabla}_p\bar{f}_{eq}\cdot\vec{\nabla}_r
(\delta{U}'_{self}+{U}'_{ext})=0.
\end{equation}
Here, $\bar{f}_{eq}$ is the equilibrium distribution function
which is the same for both protons and neutrons
according to the above-made assumptions
\begin{equation}\label{eq64}
\int{\frac{2d\vec{p}}{(2\pi\,\hbar)^3}}\,\bar{f}_{eq}(\vec{r},\vec{p})
=\rho_{n,eq}=\rho_{p,eq}=\frac{p_F^3}{3\pi^2\hbar^3}, \quad
\vec{\nabla}_p \bar{f}_{eq}(\vec{r},\vec{p})=
-\vec{v}\frac{m^*}{p_F}\delta (p-p_F).
\end{equation}
The variation of the self-consistent field $\delta{U}'_{self}$ in
Eq. (\ref{eq63}) has the form which is similar to Eq. (\ref{eq22})
\begin{equation}\label{eq65}
\delta{U}'_{self}=\int{\frac{2Vd{\vec{p}}'}{(2\pi\hbar)^3}\,{v}'_{int}
(\vec{p},{\vec{p}}')} \,\delta f(\vec{r},{\vec{p}}';t),
\end{equation}
where the effective interaction ${v}'_{int}(\vec{p},{\vec{p}}')$
for the isovector channel in the Landau approximation is given by
\cite{Mi} (see also (\ref{eq23}))
\begin{equation}\label{eq66}
{v}'_{int}(\vec{p},{\vec{p}}')=\frac{1}{N_F}\sum\limits_{l = 0}^\infty
{F}'_l P_l (\cos\theta_{p{p}'}).
\end{equation}
The interaction amplitudes ${F}'_l$ for the isovector channel
differ from the analogous ones ${F}_l$ for the isoscalar channel
in Eq. (\ref{eq23}). Thus, in contrast to the amplitude $F_0$,
which determine the nuclear compressibility modulus (see
(\ref{eq46})), the similar isovector amplitude ${F}'_0$ determines
the coefficient of isotopic symmetry $C_{sym}$ in the
Weizs\"{a}cker mass formula \cite{Mi,BoMo}
\begin{equation}\label{eq67}
C_{sym}=\frac{2}{3}\varepsilon_F \,(1+{F}'_0) \approx 60\,MeV.
\end{equation}
Below, as it was earlier done for the isoscalar channel in Eq.
(\ref{eq25}), we assume that
\begin{equation}\label{eq68}
{F}'_0\ne 0,\quad {F}'_1\ne 0,\quad F'_{l\ge 2}=0.
\end{equation}

Solving the kinetic equation (\ref{eq63}) in the same manner as
Eq. (\ref{eq21}), we find the isovector response function
${\chi}'(\omega)$ like $\chi(\omega)$ from (\ref{eq41}) as
\begin{equation}\label{eq69}
{\chi}'(\omega)=\frac{\overline{Q}_{00}(s)}{1-{\kappa}'(s)\overline{Q}_{00}(s)}
\end{equation}
where
\[
{\kappa }'(s) = - \frac{1}{N_F }\left( {{F}'_0 + \frac{{F}'_1 }{1 + {F}'_1 /
3}s^2} \right).
\]

The frequencies of isovector eigenvibrations (the poles of the
response function (\ref{eq69})) can be obtained from the
dispersion equation
\begin{equation}\label{eq70}
1-{\kappa}'(s)\overline{Q}_{00}(s)=0.
\end{equation}
The EWS $m_{-1}$, $m_1$ and $m_3$ (\ref{eq49})
for the isovector excitations take the form
\begin{equation}\label{eq71}
{m}'_{-1}=\frac{A}{2}\frac{1}{C_{sym}}, \quad {m}'_1=\hbar^2\frac{A}{2{m}'}q^2,
\quad {m}'_3=\hbar^4\frac{A}{2}\frac{{C}'_{sym}}{{m}'^2}q^4.
\end{equation}
Here, we have introduced the renormalized isotopic symmetry energy
$C'_{sym}=C_{sym}+8\varepsilon_F/15$ and the effective mass
${m}'_1$ for the isovector channel,
\[
{m}'=\frac{m}{1+\kappa_I},
\]
where $\kappa_I $ is the enhancement factor of the sum rule which
is defined by the relation
\begin{equation}\label{eq72}
1+\kappa_I=\frac{1+{F}'_1/3}{1+F_1/3}.
\end{equation}

Note that, in contrast to the isoscalar sum $m_1$ (see
(\ref{eq49})), the sum ${m}'_1$ in (\ref{eq71}) is not model
independent in sense that it depends on the effective mass ${m}'$
and thereby on the interaction amplitudes $F_1$ and ${F}'_1$. It
is worth nothing that the continuity equation (\ref{eq29})
for the isovector excitations should be modified as well. Evaluating
the zero moment from the kinetic equation (\ref{eq63}), we obtain
(see also (\ref{eq27})-(\ref{eq29}))
\begin{equation}\label{eq73}
\frac{\partial}{\partial t}\delta{\rho}'+\vec{\nabla}(1+\kappa_I)
\bar{\rho}{\vec{u}}'=0.
\end{equation}
Here
\[
\bar{\rho}=\frac{1}{2}(\rho_n+\rho_p)=\frac{\rho}{2}, \quad \quad
\delta{\rho}'=\delta\rho_n-\delta\rho_p=\int{\frac{2d\vec{p}}{(2\pi\hbar)^3}\delta{f}',}
\]
\[
{\vec{u}}'=\vec{u}_n-\vec{u}_p=\frac{1}{\bar{\rho}}\int{\frac{2
d\vec{p}}{(2\pi\hbar)^3}\frac{\vec{p}}{m}\delta{f}'\approx}
\frac{1}{\bar{\rho}_{eq}}\int{\frac{2d\vec{p}}{(2\pi\hbar)^3}\frac{\vec{p}}{m}
\delta{f}'}.
\]

Finally, the EWS (\ref{eq71}) allow one to
calculate the energy centroids of isovector giant resonances
for the adiabatic, ${\tilde{E}}'_1$, and scaling,
${\tilde{E}}'_3$, approximations as
\begin{equation}\label{eq74}
{\tilde{E}}'_1=\sqrt{\frac{{m}'_1}{{m}'_{-1}}}=\hbar\sqrt{\frac{C_{sym}}{{m}'}} \,q,
\quad {\tilde{E}}'_3=\sqrt{\frac{{m}'_3}{{m}'_1}}=\hbar\sqrt{\frac{{C}'_{sym}}{{m}'}} \,q.
\end{equation}

It is useful to compare relations (\ref{eq74}) with the
corresponding expressions (\ref{eq50}) obtained for the isoscalar
excitations.

\section{Finite nuclei. Boundary conditions}

The above-developed approach can be directly applied to the study
of the dynamic properties of the infinite nuclear matter, where the
distribution function distortion $\delta f$ has the form of a
plane wave in the $\vec{r}$-space. Below, we also apply this
approach to the description of the collective excitations in
finite nuclei. For heavy nuclei, one can assume a sharp
nuclear surface \cite{BoMo}. Then the variation $\delta f$ of the
distribution function in the nuclear interior has the form of the
plane wave (\ref{eq32}) or its projections on the states with a
fixed multipolarity. Moreover, the equation of motion must be
supplemented by the boundary conditions at the moving nuclear
surface.

To establish the boundary conditions, we introduce the force
$\vec{F}$ which is caused by a sound wave and applied to a
unit of the nuclear surface $S$, as well as the surface force $\vec{F}_S$
which is caused by a deformation of the nuclear surface. The general
condition of the equilibrium for all forces applied to the free
nuclear surface reads
\begin{equation}\label{eq75}
\vec{n}\cdot\vec{F}\vert_S+\;\vec{n}\cdot\vec{F}_S=0,
\end{equation}
where $\vec{n}$ is a unit vector in the normal direction to the
nuclear surface. Equation (\ref{eq75}) represents the boundary
condition to the dispersion equations (\ref{eq43}) and (\ref{eq70}).

To evaluate the force $\vec{F}$, we calculate the first
moment to the kinetic equation (\ref{eq21}). Multiplying Eq.
(\ref{eq21}) by $2d\vec{p}\;p_\nu /(2\pi\hbar)^3$ and integrating
over $\vec{p}$, we obtain the Eiler equation in the following
form \cite{Ko,KoSh,KoMaPl}
\begin{equation}\label{eq76}
\rho_{eq}\frac{\partial}{\partial t}u_\nu=-\nabla_\mu\delta\Pi_{\nu\mu},
\end{equation}
where $\delta\Pi_{\nu\mu}$ is the pressure tensor. For the
isovector excitations, we obtain \cite{KoSh}
\begin{equation}\label{eq77}
\delta\Pi_{\nu\mu}=\delta{\sigma}'_{\nu\mu}+\delta{P}'\,\delta_{\nu\mu},
\end{equation}
where
\[
\delta {P}'\equiv\delta{P}'\left({\vec{r},t}\right)
=\frac{1}{3m}\int
{\frac{2d\vec{p}}{\left({2\pi\hbar}\right)^3}}
p^2\delta{f}'\left({\vec{r},\vec{p},t}\right)+\frac{1}{N_F}{F}'_0\bar{\rho}_{eq}
\delta\rho'\left({\vec{r},t}\right)=
\]
\begin{equation}\label{eq78}
=\frac{2}{3}\left({1+{F}'_0}\right)\varepsilon_F\delta{\rho}'
\left({\vec{r},t}\right)=C_{sym}\delta{\rho}'\left({\vec{r},t}\right).
\end{equation}
Let us introduce the isovector displacement field ${\vec{\chi}}'$
which is connected to the corresponding velocity field
${\vec{u}}'$ through the relation
\[
\partial{\vec{\chi}}'/\partial t=-(1+\kappa_I)\vec{u}'.
\]
Taking the continuity equation (\ref{eq73}) into account, we find
\[
\delta{\rho}'=\bar{\rho}_{eq}\vec{\nabla}\cdot{\vec{\chi}}'.
\]
Finally, Eq. (\ref{eq78}) yields
\begin{equation}\label{eq79}
\delta{P}'=C_{sym}\bar{\rho}_{eq}\vec{\nabla}\cdot\vec{{\chi}'}.
\end{equation}
The pressure tensor $\delta{\sigma}'_{\nu\mu}$ in Eq. (\ref{eq77})
is given by
\begin{equation}\label{eq80}
\delta{\sigma}'_{\nu\mu}=\frac{2}{3m^\ast}\int{\frac{d\vec{p}}{(2\pi\hbar)^3}}
(3p_\nu p_\mu-p^2)\delta{f}' {\mu}'_F(\nabla_\nu{\chi}'_\mu +\nabla_\mu{\chi}'_\nu -
\frac{2}{3}\ \delta_{\nu\mu}\vec{\nabla}\cdot{\vec{\chi}}'),
\end{equation}
where
\begin{equation}\label{eq81}
{\mu}'_F=\frac{3}{2}\ \bar{\rho}_{eq}\varepsilon_F\ \frac{s^2}{1+{F}'_1/3}
\left[{1-\frac{(1+{F}'_0)(1+{F}'_1/3)}{3s^2}}\right].
\end{equation}
Taking (\ref{eq77}), (\ref{eq79}) and (\ref{eq80}) into account,
we obtain
\begin{equation}\label{eq82}
\delta\Pi_{\alpha\beta}={\mu}'_F\left({\nabla_\alpha{\chi}'_\beta
+\nabla_\beta{\chi}'_\alpha}\right)
+\left({C_{sym}\bar{\rho}_{eq}-\frac{2}{3}\ {\mu}'_F}\right)\vec{\nabla}
\cdot{\vec{\chi}}'\delta_{\alpha\beta }
\end{equation}
The pressure tensor $\delta\Pi_{\nu\mu}$ determines the force
$\vec{F}$ which acts from the side of the sound wave on a unit of the
nuclear surface
\begin{equation}\label{eq83}
F_\nu=n_\mu\delta\Pi_{\nu\mu}.
\end{equation}
Using Eqs. (\ref{eq82}) and (\ref{eq83}), we evaluate the normal
component of the force $\vec{F}$ applied to the nuclear surface:
\[
\vec{n}\cdot\vec{F}\left\vert_S=\frac{1}{r^2}r_\nu r_\mu\delta\Pi_{\nu\mu}\right\vert_{r=R_0}
=\frac{1}{r^2}\left[r^2\left(C_{sym}\bar{\rho}_{eq}(1+\kappa_I)-\frac{2}{3}{\mu}'_F \right)
\vec{\nabla}\cdot{\vec{\chi}}'+ 2 {\mu}'_F\ r_\nu r_\mu\chi_\nu\chi_\mu\right]_{r=R_0}
\]
\begin{equation}\label{eqq}
=\left[\left({C_{sym}\bar{\rho}_{eq}(1+\kappa_I)-\frac{2}{3}\
{\mu}'_F}\right)\mbox{div}{\vec{\chi}}'
+2{\mu}'_F\frac{\partial}{\partial r}\left({\vec{n}\cdot{\vec{\chi}}'}
\right)\right]_{r=R_0}.
\end{equation}

Let us calculate the normal component $\vec{n}\cdot\vec{F}_S$ of
the isovector surface force $\vec{F}_S$ which occurs in Eq.
(\ref{eq75}). To find the force $\vec{F}_S$, we notice that a
shift of protons against neutrons creates the additional surface
energy in the case of isotopic symmetry given by \cite{MySw}
\begin{equation}\label{eq84}
\delta E_{S,sym}=\frac{1}{3}\rho_{eq} r_0\sigma_{sym}\int{\tau ^2dS}.
\end{equation}
Here, $r_0$ is the mean distance between nucleons
($R_0=r_0A^{1/3})$, $\sigma_{sym}$ is the isovector surface
energy which is a parameter of theory, and $\tau$ is a shift of
the proton sphere against the neutron one. In units of $r_0$,
\begin{equation}\label{eq85}
\tau=\frac{1}{r_0}\left({R_p\left(t\right)-R_n\left(t\right)}\right)
=\frac{1}{r_0}\left({\left({R_0+\delta R_1(t)}\right)-\left({R_0-\delta
R_1(t)}\right)}\right)=\frac{2}{r_0}\delta R_1(t),
\end{equation}
where
\begin{equation}\label{eq86}
\delta R_1(t)=R_0\alpha_S\left(t\right)Y_{10}\left(\hat{r}\right).
\end{equation}
The amplitude $\alpha_S(t)$ of isovector vibrations of the nuclear
surface in Eq. (\ref{eq86}) is connected to the corresponding
amplitude ${\vec{\chi}}'$ of the displacement field in a sound wave.
To establish this connection, we note that, for a nucleus with
the sharp edge, the displacement field in nuclear interior has the
form (see Section 6 in \cite{BoMo})
\begin{equation}\label{eq87}
{\vec{\chi}}'=\alpha_1\left(t\right)\frac{1}{q^2}\vec{\nabla}_r
\left({j_1\left({qr}\right)Y_{10}\left(\hat{r}\right)}\right),
\end{equation}
where $j_1(x)$ is the spherical Bessel function.

Evaluating the normal component of the velocity field ${\vec{u}}'$
by the use of Eq. (\ref{eq87}) and equating it to the surface
velocity $\partial\delta R_1(t)/\partial t$, we obtain
\begin{equation}\label{eq88}
\alpha_S\left(t\right)=-\alpha_1\left(t\right)\frac{{j}'_1\left(x\right)}
{x(1+\kappa_I)}, \quad x=qR_0.
\end{equation}
According to the definition of the pressure $\delta P_S$ caused
by a shift of the nuclear surface (see, for example, the appendix
to Section 6 in \cite{BoMo} ), we obtain the following relation
from Eqs. (\ref{eq84}) and (\ref{eq85}):
\begin{equation}\label{eq89}
\delta P_S=\frac{\partial}{\partial\delta R_1}\frac{\delta E_S}{\delta S}
=\frac{8}{3}\frac{\rho_{eq}}{r_0}\ \sigma_{sym}\delta R_1 .
\end{equation}
Taking into account Eqs. (\ref{eq88}) and  (\ref{eq89}), we can
evaluate the normal component ($\vec{n}\cdot\vec{F}_S$) of the
surface force $\vec{F}_S$ in Eq. (\ref{eq75}). The result reads
\begin{equation}\label{eq90}
\vec{n}\cdot\vec{F}_S=-\delta P_S=\frac{8}{3}\frac{\rho_{eq}{j}'_1 (x)}
{qr_0(1+\kappa_I)}\ \sigma_{sym}\alpha _1 \left(t\right)Y_{10}\left(\hat{r}
\right).
\end{equation}
Finally, from Eqs. (\ref{eq75}), (\ref{eqq}), (\ref{eq87}) and
(\ref{eq90}) we derive the following secular equation for the wave
number $q$:
\begin{equation}\label{eq91}
\left[{-\frac{1}{2}C_{sym}\bar{\rho}_{eq}-\frac{2}{3}{\mu }'_F
+ \frac{2}{x^2}{\mu}'_F}\right]j_1 \left(x\right)
+\left[{-\frac{2}{x}{\mu }'_F+\frac{4}{3}\frac{\rho_{eq}}{qr_0(1+\kappa_I)}
\sigma_{sym}}\right]{j}'_1\left(x\right)= 0.
\end{equation}
We point out that in the classical limit of the
Steinwedel-Jensen's model at $\sigma_{sym}\to\infty$, the boundary
condition (\ref{eq91}) coincides with the similar one,
${j}'_1(x)=0$, in the traditional liquid drop model \cite{BoMo}.
\begin{figure}
\includegraphics{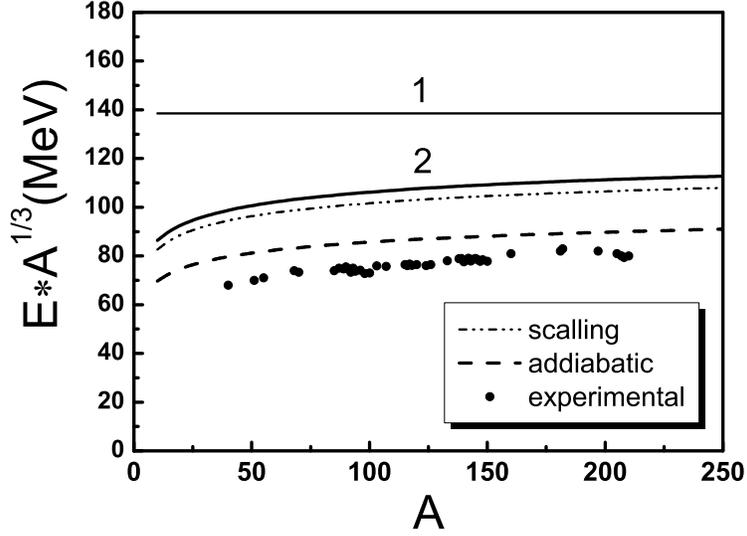}
\caption{Dependence of the energy of the isovector giant dipole resonances
on the mass number obtained from the dispersion equation (\ref{eq70}) (solid
curve 2) and from EWS (\ref{eq74}) (dashed lines). Solid curve 1 is obtained
from the explicit solution of the dispersion equation (\ref{eq70}) subsidized
by the boundary condition of the Steinwedel-Jensen's model, ${j}'_1(x)=0$. For
all calculations presented in Fig. 2, we have taken the following parameters:
$r_0=1.2$ fm, $F_1=-0.64$, ${F}'_0=0.96$, ${F}'_1 =1$, $\sigma_{sym}=17$ MeV.
The experimental data were taken from \cite{BeFu}.}
\label{fig2}
\end{figure}

The boundary condition (\ref{eq91}) allows us to find the
dependence of the wave number $q$ on the mass number $A$ and to evaluate
the corresponding excitation energy in finite nuclei. In \figurename\
\ref{fig2}, we show the dependence of the energy of isovector giant dipole
resonances (IGDR) on the mass number $A$ obtained by the use of the
explicit solution of the dispersion equation (\ref{eq70}) and
EWS (\ref{eq74}). For both of them, the boundary condition
(\ref{eq91}) was used. As can be seen from \figurename\  \ref{fig2},
the lowest energy of IGDR $\sqrt{m'_1/m'_{-1}}$ is obtained
with $m'_{-1}$ and corresponds to the first sound regime without
the Fermi surface distortions. The account of a quadrupole Fermi
surface deformation in the sum $m'_3$ shifts upward the
curve $\sqrt{m'_3/m'_1}$. This is due to the additional
contribution to the nuclear stiffness coefficient caused by
Fermi-surface distortions. Involving the higher multipolarities
of the Fermi surface distortions which are present in the
dispersion equation (\ref{eq70}) leads to the additional increase
of the nuclear stiffness and the excitation energy
$\hbar\omega_{1^-}$ in \figurename\  \ref{fig2}.

\begin{figure}
\includegraphics{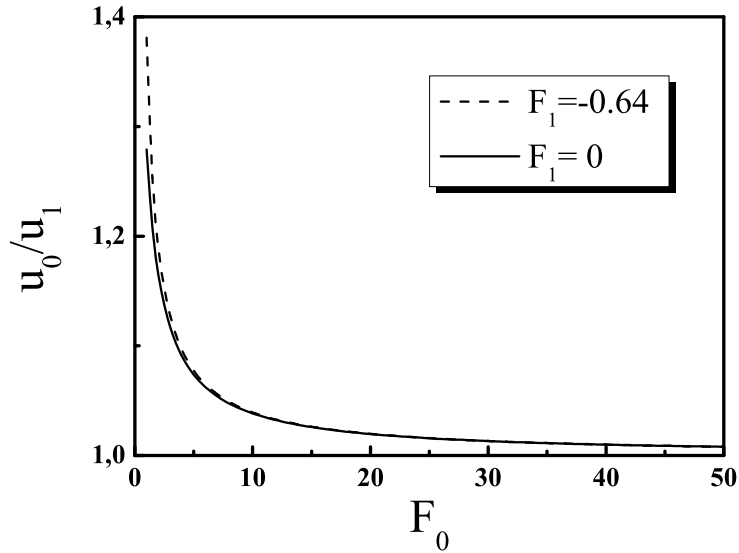}
\caption{Dependences of the ratio of the first sound velocity to the
zero-sound one on the interaction amplitude $F_0$ for the isoscalar
excitations for two values of the interaction constant $F_1$.}
\label{fig3}
\end{figure}

\begin{figure}
\begin{center}
\includegraphics{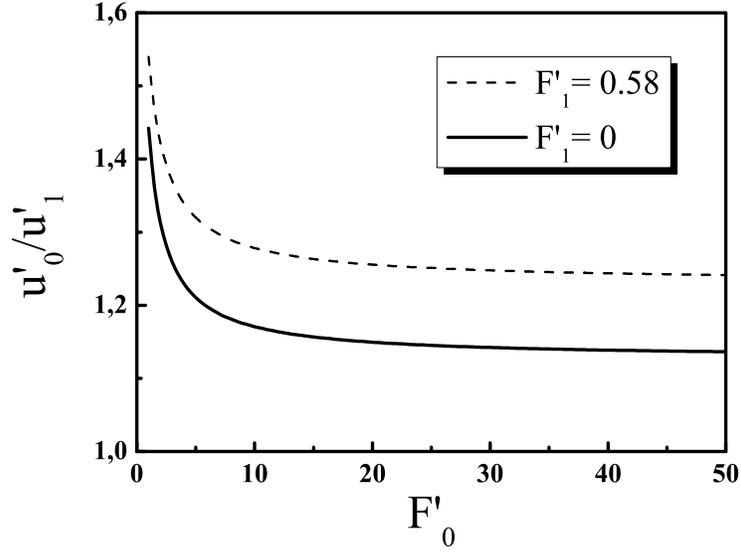}
\end{center}
\caption{The same as in \figurename\  \ref{fig3}, but for isovector
excitations.}
\label{fig4}
\end{figure}

As was noted above, the dependence of the nuclear forces on the
nucleon velocities (components with $F_1$ and ${F}'_1$ in
(\ref{eq23}) and (\ref{eq66}), respectively) leads to the
significant difference between the EWS for the
isoscalar and isovector excitations. In particular, the
consequence of this difference is the asymptotic behavior of the
nuclear stiffness coefficient and the zero-sound velocity at an
increase of the internucleon interactions $F_0$ and ${F}'_0$.
In \figurename\ \ref{fig3} and \ref{fig4}, we show the dependence
of the ratio of the zero-sound velocity to the first sound one on
the interaction amplitudes $F_0$ and ${F}'_0$ for the isoscalar
and isovector excitations. The feature of the isoscalar excitations
is the fact that the increase of the nucleon-nucleon interaction
leads to a shift of the zero-sound velocity towards the first
sound one (see \figurename\ \ref{fig3}). This means that the influence
of Fermi surface distortions on the collective motion in the nuclear
Fermi-liquid becomes negligible on the increase of the nucleon-nucleon
interaction. The behavior of the isovector zero-sound velocity
${u}'_0$ is qualitatively different (see \figurename\  \ref{fig4}).
With increase in the  nucleon-nucleon interaction, the velocity ${u}'_0$
tends to the asymptotic limit which significantly exceeds the
corresponding first sound velocity. This is a consequence of the
general enhancement effect of collectivity of the isoscalar
zero-sound caused by the dependence of the nuclear forces on the
nucleon velocity (see also (\ref{eq72})).

The enhancement factor $\kappa_I$ for the isovector excitations
defined in Eq. (\ref{eq72}) depends on the interaction constants
$F_1$ and ${F}'_1 $.  Whereas the isoscalar constant $F_1$
related to the effective nucleon mass $m^*$ is well studied,
the isovector constant ${F}'_1$ is not much studied, and the
experimental investigation of the enhancement factor $\kappa_I$
(\ref{eq72}) can help for its derivation.

The experimental derivation of the enhancement factor in the
isovector EWS ${m}'_1$ is connected to the investigation of the
nuclear absorption cross-section $\sigma_{abs}(\omega)$ of
$\gamma$-quanta with energy $\hbar\omega$. Let us introduce the
strength function $S(\omega,q)$ for the density-density response
per unit volume $V$. According to Eqs. (\ref{eq1}), (\ref{eq40}),
and (\ref{eq42}), we have
\begin{equation}\label{eq92}
S(\omega,q)=\frac{1}{V}\chi^{(d)}(\omega).
\end{equation}
In the case of the velocity independent forces, in accordance with
Eqs. (\ref{eq2}), (\ref{eq49}) and (\ref{eq71}), the strength
function $S(\omega,q)$ is normalized by the condition
\begin{equation}\label{eq93}
\int\limits_0^\infty{d(\hbar\omega)\,\hbar\omega S}(\omega,q)=
\hbar^2\frac{1}{2m}\,q^2\rho_{eq}.
\end{equation}
In contrast to this, in the case of isovector excitations with the
velocity dependent forces, the normalization condition reads
\begin{equation}\label{eq94}
\int\limits_0^\infty{d(\hbar\omega)\,\hbar\omega S}(\omega,q)=\hbar
^2\frac{A}{2m^*}(1+{F}'_1/3)\,q^2\rho_{eq}.
\end{equation}
The photoabsorption cross-section $\sigma_{abs}(\omega)$ is
connected to the strength function by the relation
\begin{equation}\label{eq95}
\sigma_{abs}(\omega)=const\cdot\omega S(\omega,q).
\end{equation}
Here, $\vec{q}$ plays the role of the momentum which is transferred to
the nucleus at the absorbtion of a $\gamma$-quantum. The constant
in Eq. (\ref{eq95}) can be found from the normalization condition
for the photoabsorption cross-section $\sigma_{abs}(\omega)$. For
the isovector dipole excitations and for the velocity independent
forces, this condition reads (Reiche-Thomas-Kuhn rule) \cite{RiSh}
\begin{equation}\label{eq96}
\tilde{m}_1=\int\limits_0^\infty {d(\hbar\omega)\;\sigma_{abs}(\omega)}
=\frac{2\pi^2\hbar e^2}{mc}\frac{NZ}{A}.
\end{equation}
From Eqs. (\ref{eq95}), (\ref{eq93}) and (\ref{eq96}), one obtains
\cite{ToKoLa}
\begin{equation}\label{eq97}
\sigma_{abs}(\omega)=\frac{4\pi^2e^2}{cq^2\rho_{eq}}
\frac{NZ}{A}\omega S(\omega,q).
\end{equation}
Since the photoabsorption occurs mainly through the giant dipole
resonance, the transferred momentum $q$ in (\ref{eq97}) can be taken
as $q=q_1=2.08/R_0$ \cite{ToKoLa}, that corresponds to the
classical boundary condition ${j}'_1(x)=0$ of the
Steinwedel-Jensen's model.

\begin{figure}
\noindent\includegraphics{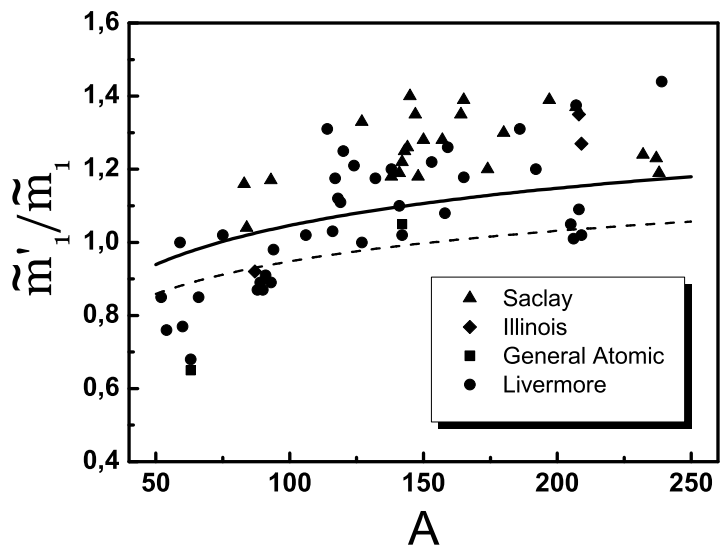}
\caption{Dependence of the enhancement factor ${\tilde {m}}'_1/\tilde{m}_1$
of the EWS ${m}'_1$ for isovector giant dipole resonances on the mass
number $A$. The results of calculations were obtained by
the use of Eqs. (\ref{eq96}) and (\ref{eq98}) for two values of
the isovector amplitude ${F}'_1=1.2$ (solid line) and
${F}'_1=0.58$ (dashed line) at the constant value of $F_1=-0.64$;
points are the experimental data from \cite{Bo}.}
\label{fig5}
\end{figure}

In the case of the velocity dependent forces, the normalization
condition $S(\omega,q)$ (\ref{eq93}) has to be replaced by
condition (\ref{eq94}), and the transferred momentum $q={q}'_1$ has to
be calculated by using the boundary condition (\ref{eq91}). As
a result, the sum rule for $\sigma_{abs}(\omega)$ takes the form
(instead of (\ref{eq96}))
\begin{equation}\label{eq98}
{\tilde{m}}'_1=\int\limits_0^\infty{d(\hbar\omega)\sigma_{abs}(\omega)}
=\frac{2\pi^2\hbar e^2}{mc}\frac{NZ}{A}\left({\frac{{q}'_1}{q_1}}\right)^2
(1+\kappa_I).
\end{equation}
In \figurename\  \ref{fig5}, we demonstrate the dependence of the
enhancement factor ${\tilde{m}}'_1/\tilde{m}_1$ on the mass number for
a number of nuclei.
We point out that the exceeding of 100{\%} of the
sum rule ${\tilde{m}}'_1$, which is experimentally observed for the
isovector giant dipole resonances, is caused by the dependence of
the effective nucleon-nucleon interaction on the nucleon velocity.
For the value of the isovector amplitude ${F}'_1\approx 1$,
one can adjust (on the average) the results of theoretical calculations
of ${\tilde{m}}'_1$ (solid line in \figurename\ \ref{fig5}) with the
experimental data. The nonmonotonic dependence of the experimental value
of ${\tilde{m}}'_1$ on the mass number $A$ in \figurename\ \ref{fig5}
is due to the shell effects which are not taken into account within
the semiclassical kinetic theory used in this work.

\section{Conclusions}

Using the Landau-Vlasov kinetic theory, we have studied the linear
response function and the EWS $m_k$ for the
isoscalar and isovector excitations in heavy nuclei and the nuclear
matter. An advantage of our approach is the possibility to derive
the explicit analytical expressions and to carry out a detailed
analysis for some important nuclear characteristics. One of the
them is the nuclear stiffness coefficient. The dynamical Fermi-
surface distortion influences significantly the formation of the
nuclear stiffness. For a slow (adiabatic) nuclear deformation, the
nuclear stiffness coefficient is derived by the low-energy sum
$m_{-1}$ and coincides with the stiffness coefficient of the
classical (non-Fermi) liquid. This stiffness coefficient (adiabatic
incompressibility $K$) causes the propagation of the first sound
in a Fermi-liquid which is not accompanied by the Fermi surface
distortions [see (\ref{eq10}), (\ref{eq47}), and (\ref{eq49})]. In
the general case of fast motion, the derivation of the nuclear
stiffness coefficient requires to solve the dispersion equation
(\ref{eq43}). A specific role is played here by the scaling
approximation. In this last case, the quadrupole Fermi surface
distortion is only taken into consideration. We have shown that
the stiffness  coefficient in the scaling approximation is
determined by the high-energy sum $m_3$ and exceeds significantly
the adiabatic incompressibility $K$. At the same time, the sound
velocity approaches that of the Landau's first sound [see
(\ref{eq48}), (\ref{eq51}), and (\ref{eq59})].

In the presence of the velocity dependent nuclear forces, the
EWS for the isoscalar and isovector excitations
are significantly different [see (\ref{eq49}) and (\ref{eq71})].
First of all, the EWS $m_1$, which is model
independent for the isoscalar excitations, becomes model dependent
in the case of the isovector excitations. Another consequence of
the mentioned difference of sums (\ref{eq49}) and (\ref{eq71}) is
the different asymptotic behavior of the zero-sound velocity on an
increase of the nucleon-nucleon interaction (see \figurename\  \ref{fig3}
and \ref{fig4}).

A feature of the isoscalar excitations is that the zero-sound
velocity approaches the first sound one with increase in the
internucleon interaction. This means that the influence of the
Fermi surface distortion on the isoscalar collective motion in
the nuclear Fermi-liquid becomes negligible on the increase of
the internucleon interaction. In contrast to this, the increase
of the internucleon interaction for the isovector mode leads to
the asymptotic zero-sound velocity ${u}'_0$ which exceeds the
relevant first sound velocity. The above-mentioned difference
between isoscalar and isovector EWS allows one to explain the
fact that, in many cases, the experimental measurement of the
EWS ${m}'_1$ for the isovector giant dipole resonance gives the
more than 100 {\%} exhaustion of the corresponding sum rule.
According to Eqs. (\ref{eq71}) and (\ref{eq72}), the dependence
of the effective nuclear forces on the nucleon velocities
generates the enhancement factor $1+\kappa_I>1$, which is absent
for the isoscalar excitations, in the sum ${m}'_1$ for the
isovector excitations (see \figurename\  \ref{fig5}). Note that
the enhancement factor (\ref{eq72}) depends on the isovector
amplitude ${F}'_1$. This gives, in principe, the possibility
to determine the interaction amplitude ${F}'_1$ from the fit
of the EWS ${\tilde{m}}'_1$ to the experimental data.


\begin{thebibliography}{99}

\bibitem{BoLaMa} O. Bohigas, A.M. Lane and J. Martorell,
Phys. Rep. {\bf 51}, 267 (1979).

\bibitem{La} A.M. Lane,
Description of giant resonances, in:
Proc. Intern. Symp. on Highly excited states in nuclei. J\"{u}lich
(1975) p. 95.

\bibitem{LiBr} K.F. Liu and G.E. Brown,
Nucl. Phys. {\bf A 265}, 385 (1976).

\bibitem{HaDi} M.N. Harakeh and A.E.L. Dieperink,
Phys. Rev. {\bf C 23}, 2329 (1981).

\bibitem{St} S. Stringari, Phys. Lett. {\bf B 108}, 232 (1982).

\bibitem{Wo} A. Van der Woude, The electric giant resonances, Preprint KVI-820,
Groningen (1989).

\bibitem{LiSt} E. Liparini and S. Stringari, Phys. Rep. {\bf 175}, 103 (1989).

\bibitem{HaSaZh97} I. Hamomoto, H. Sagawa and X.Z. Zhang, Phys. Rev. {\bf C 56},
3121 (1997).

\bibitem{HaSaZh98} I. Hamomoto, H. Sagawa and X.Z. Zhang, Phys. Rev. {\bf C 57},
R1064 (1998).

\bibitem{SaHaZh99} H. Sagawa, I. Hamomoto and X.Z. Zhang, Nucl.
Phys. {\bf A 649}, 319c (1999).

\bibitem{Ko} V.M. Kolomietz, \textit{Local density approximation in atomic and
nuclear physics} (Naukova Dumka, Kyiv, 1990).

\bibitem{KoSh} V.M. Kolomietz and S. Shlomo, Phys. Rep. {\bf 390}, 133 (2004).

\bibitem{LaLi5} E.M. Lifshitz, L.P. Pitaevsky,  \textit{Statistical physics}, Pt 1,
(Nauka, Moscow, 1976).

\bibitem{AbKha} A.A. Abrikosov, I.M. Khalatnikov, Rep. Prog. Phys. {\bf 22}, 329 (1959).

\bibitem{BaPe} G. Baym and C.J. Pethick, \textit{Landau Fermi liquid theory} (J.
Wiley {\&} Sons, New York, 1991).

\bibitem{LaLi10} E.M. Lifshitz, L.P. Pitaevsky \textit{Physical kinetics}, (Nauka,
Moscow, 1978).

\bibitem{Mi} A.B. Migdal, \textit{Theory of finite Fermi systems and applications to
atomic nuclei} (Interscience, London, 1967).

\bibitem{BoMo} A. Bohr, B. Mottelson, \textit{Nuclear structure}, Vol.2 (W.A. Benjamin,
New York, 1975).

\bibitem{KoKoSh} A. Kolomiets, V.M. Kolomietz and S. Shlomo, Phys. Rev.
{\bf C 59}, 3139 (1999).

\bibitem{Bo} O. Bohigas, Suppl. Prog. Theor. Phys. {\bf 74 -75}, 380 (1983).

\bibitem{KoMaPl} V.M. Kolomietz, A.G. Magner and V.A. Plujko, Z. f\"{u}r Phys.
{\bf A 345}, 137 (1993).

\bibitem{MySw} W.D. Myers and W.J. Swiatecki, Ann. Phys. {\bf 84}, 186 (1974).

\bibitem{RiSh} P. Ring and P. Schuck, \textit{The nuclear many-body problem},
(Springer-Verlag, New-York, 1980).

\bibitem{ToKoLa} M. Di Toro, V.M. Kolomietz and A.B. Larionov, Phys. Rev.
{\bf C 59}, 3099 (1999).

\bibitem{BeFu} B.L. Berman and S.C. Fultz, Rev. Mod. Phys. {\bf 47}, 713
(1975).

\end{thebibliography}
\end{document}